# The Core Diffusion-Drift Field-Effect Transistor Theory Including Quantum and Interface Trap Capacitances

Gennady I. Zebrev

***Abstract*** — We have decomposed the modeling of the field-effect transistors into the two independent parts: the current continuity based kinetics and the charge neutrality based electrostatics. The former part, that is universal for all FETs, leads to an explicit and closed form of I-V characteristics as a function of the total channel charge. The latter part, which is specific for a particular material and geometric configurations can be considered as an independent engineering task. The quantum capacitance and the interface trap density are consistently incorporated into the solution of the current continuity equation in the diffusion-drift approximation, providing a complete consistency in the current and the capacitance small-signal characteristics.

## I. INTRODUCTION

The demands of advanced chip design in microelectronics is pushing the industry to develop new compact MOSFET models for design of advanced chip architectures [1]. A comparative analysis of different approaches to the MOSFET compact modeling was given in [2]. It is well-known the channel current in the field-effect transistors has in principle a diffusion-drift form due to fundamental non-uniformity of the charge distribution along the channel [3]. Despite this fact, most of the compact MOSFET models based anyway on a formal integration of the drift-like dependence $I_D = G\, dV/dx$ [4, 5], where $G$ is the channel conductance formally depending on the gate $V_G$ and the drain $V_D$ biases. This inconsistency leads often to a piecewise description of transistor operation in different modes, which is artificially corrected by formal interpolation procedures. We pointed in [6], that a consistent description of I-V characteristics in nonlinear electronic devices can be constructed on only the joint solution of the Poisson and current continuity equations. We argued that such theory, relying only upon the general principles of current continuity and electric neutrality along the channel (used instead of the exact solution of Poisson's equation), is universal and can be used for any configurations of the field-effect transistors (FETs). A particular form of device geometry and material equations has to enter the theory only through the total node capacitances and through a dependence of the total channel charge $Q_C$ on the external node biases. This point is very important since a use of the areal capacitance and the local charge density variables in the deeply scaled devices is problematic because of the fringing effects and non-uniformities in nanoscale devices. We consider a calculation of the full channel charge $Q_C$ as a separate technical task, specific for different FETs. A particular form of $Q_C(V_G)$ should be determined from electrostatic consideration, which is different for the different device configurations (bulk or SOI FETs, FinFETs, double gate FETs, etc.). The fundamental quantum capacitance and the parasitic interface trap capacitance should be consistently implemented into the structure of the core model.

We describe in this paper an advanced version of the diffusion-drift model [6] meeting these requirements and formulated in a general, consistent and physically transparent form.

## II. DIFFUSION-DRIFT THEORY

### A. Diffusion to drift current ratio

The diffusion-drift nature of the channel current implies a use of the electrochemical potential notion. The full diffusion-drift current is determined by the gradient of electrochemical potential $\mu = \zeta - q\varphi$

$$J = j_{DR} + j_{DIFF} = -\mu_0 n_C \frac{d\mu}{dx} = \\ = -q\mu_0 n_C \frac{d\varphi}{dx}\left(1 - \frac{d\zeta}{qd\varphi}\right) = q\mu_0 n_C E(1+\kappa), \quad (1)$$

where $\mu_0$ is the carrier's mobility, $n_C$ is the channel electron density, $E = -d\varphi/dx$ is a lateral electric field. As shown in Ref. [6], the diffusion to the drift current ratio $\kappa$ can be derived from the electric neutrality requirement along the channel of MOS structure in the gradual channel approximation

$$\kappa \equiv \frac{j_{DIFF}}{j_{DR}} = -\left(\frac{\partial \zeta}{q\partial \varphi}\right)_{V_G} = \frac{(\partial V_G/\partial \varphi)_\zeta}{q(\partial V_G/\partial \zeta)_\varphi} = \frac{C_{OX} + C_D}{C_Q + C_{IT}}, \quad (2)$$

where $C_{OX}$, $C_D$, $C_{IT}$ are the total oxide (geometrical) capacitance, the total depletion layer and the interface trap capacitances, respectively. For convenience, we will denote the total charges and capacitances by the indices with the capital letters. Notice, that only $C_{OX}$ and $C_Q$ play a fundamental role in the field-effect device operation principle, whereas the interface trap and the depletion layer effects could be eliminated or minimized for ideal, or some particular device configurations.

The total inversion layer capacitance (referred often as a quantum capacitance [7]) $C_Q = q dQ_{C0}/d\zeta$ is closely related with the diffusion potential which is defined as a fundamental parameter of the generalized Einstein relation

$$\varphi_D \equiv \frac{Q_{C0}}{q dQ_{C0}/d\zeta} \equiv \frac{Q_{C0}}{C_Q} = \frac{D_0}{\mu_0}, \quad (3)$$

where $D_0$ is the electron diffusivity, $Q_{C0}$ is the total "equilibrium" channel charge at zero drain bias, which will be considered below as a known function of gate voltage. As well as the channel charge $Q_{C0}$, the quantum capacitance $C_Q$ varies over a very wide range, while the diffusion potential $\varphi_D$ varies only from the thermal potential $\varphi_T = k_B T/q$ in the subthreshold region to a few $\varphi_T$ (of order the Fermi energy) in the strong inversion.

The quantum capacitance is usually ignored in the MOSFET modeling. Actually, the quantum capacitance $C_Q$ is very low in the subthreshold operation mode ( $C_Q \ll C_{OX}, C_{IT}, C_D$ ) and extremely high in the strong inversion regime ( $C_Q \gg C_{OX}, C_{IT}, C_D$ ). In the former case, $C_Q$ is masked by the parasitic interface trap and the depletion layer capacitances. In the latter case, $C_Q$ is typically insignificant in the silicon FETs (in contrast to graphene FETs) due to the series connection with the geometrical capacitance $C_{OX}$ (see, the inset in Fig. 1). Nevertheless, we will see below that the quantum capacitance is a key concept in the construction of the consistent MOSFET theory.

### B. I-V characteristics from the continuity equation solution

The key point of the original approach [6] is an explicit analytical solution of the continuity equation for the channel current density. The total diffusion-drift drain current $J_S = J_{DR} + J_{IFF} \equiv (1+\kappa) J_{DR}$ has to be conserved along the channel in a quasi-static approximation. Assuming that the carrier's mobility and $\kappa$ are coordinate-independent along the channel, we have the continuity equation in a form

$$-\frac{d}{dx}\left(n_C \frac{d\mu}{dx}\right) = (1+\kappa)\frac{d}{dx}(n_C E) = 0, \quad (4)$$

that yields the equation for the electric field distribution along the channel

$$\frac{dE}{dx} = \frac{\kappa}{\varphi_D} E^2. \quad (5)$$

Notice that this equation and electric field $E(x)$ have nothing to do with the built-in electric fields in generally non-uniform channels that can be calculated as a result of the Poisson's equation even for zero drain-source bias. A direct solution of the ordinary differential equation Eq. 5 yields

$$E(x) = \frac{E(0)}{1 - \kappa E(0) x/\varphi_D}, \quad (6)$$

where $E(0)$ is the electric field near the source, which should be determined from the condition imposed by a fixed source - drain electrochemical potential difference $V_D$

$$V_D = \int_0^L \left(-\frac{d\mu(x)}{q dx}\right) dx = (1+\kappa)\int_0^L E(x) dx, \quad (7)$$

where $L$ is the channel length.

Based on (6) and (7), the explicit expression for $E(0)$ and electrochemical potential distribution along the channel can be derived

$$\mu(x) - \mu(0) = \varphi_D \frac{1+\kappa}{\kappa} \ln\left[1 - \frac{x}{L}\left[1 - \exp\left(-\frac{\kappa}{1+\kappa}\frac{V_D}{\varphi_D}\right)\right]\right]. \quad (8)$$

Here, $\mu(0)$ the electrochemical potential (qasi-Fermi level) near the source ( $\mu(L) - \mu(0) = -qV_D$ ), controlled by the gate-source bias $V_G$. The chemical potential $\zeta$ and the electron density distributions along the channel can be also inferred in this way. According to a general rule, the total diffusion-drift drain current can be calculated as a gradient of the electrochemical potential taken in a suitable point, e.g., in the vicinity of the source [8]

$$I_D = -W\mu_0 n_C(0)\left(\frac{d\mu}{dx}\right)_{x=0} = -\frac{\mu_0 Q_{C0}}{L}\left(\frac{d\mu}{q dx}\right)_{x=0}, \quad (9)$$

where $W$ is the FET width, $n_C(0)$ ( $Q_{C0} = q n_C(0) W L$ ) is electron density near the source.

Thus, the MOSFET drain current $I_D$ can be figured out in a closed explicit form as follows

$$I_D = \frac{D_0 Q_{C0}}{L^2}\frac{1+\kappa}{\kappa}\left(1 - \exp\left(-\frac{\kappa}{1+\kappa}\frac{V_D}{\varphi_D}\right)\right). \quad (10)$$

Defining the electrostatic saturation drain bias as $V_{DSAT} \equiv 2\varphi_D(1+\kappa)/\kappa$, we have

$$I_D = \frac{1}{2}G_0 V_{DSAT}\left(1 - \exp\left(-2\frac{V_D}{V_{DSAT}}\right)\right), \quad (11)$$

where the channel conductance in the linear region is defined as follows

$$G_0 \equiv \mu_0 Q_{C0}/L^2. \quad (12)$$

In view of (2) and (3), the direct calculation yields

$$V_{DSAT} = 2\varphi_D\left(1 + \frac{C_{IT}}{C_{OX} + C_D}\right) + \frac{2Q_C}{C_{OX} + C_D}. \quad (13)$$

We took into account in (13) that the electrostatic saturation regime corresponds to a non-uniform charge distribution in the channel. In contrast to $Q_{C0}$, the total "nonequilibrium" channel charge $Q_C = Q_C(V_G, V_D)$ should also depend on the drain-source bias.

The formula (14) yields a convenient approximation for the self-consistent calculation of $Q_C(V_G, V_D)$ provided $Q_{C0}$ is known

$$Q_C(V_G, V_D) = \frac{Q_{C0}}{2}\left(1 + \exp\left(-\frac{\kappa}{1+\kappa}\frac{V_D}{\varphi_D}\right)\right) \cong$$
$$\cong \frac{Q_{C0}}{2}\left(1 + \exp\left(-\frac{(C_{OX}+C_D)V_D}{Q_{C0}}\right)\right). \quad (14)$$

Note, that $Q_C(V_G, V_D = 0) = Q_{C0}$, $Q_C(V_G, V_D > V_{DSAT}) \cong Q_{C0}/2$, and $V_{DSAT} \cong Q_{C0}/(C_{OX}+C_D)$ in strong inversion. Thus, the equations (11-14) represent a closed, consistent and explicit analytical computational scheme for evaluation of the drain current as a function of $Q_{C0}$.

It is notable that I-V characteristics of the field-effect transistors can be equivalently and self-consistently rewritten in an intuitively clear and concise form

$$I_D = \frac{Q_C}{\tau_{TT}}, \quad (15)$$

where the transit time is defined as

$$\tau_{TT} = \frac{L^2}{\mu_0 V_{DSAT}} \coth\left(\frac{V_D}{V_{DSAT}}\right). \quad (16)$$

The same relation for the transit time can be consistently obtained in the following way [9]

$$\tau_{TT} = \int_0^L \frac{dx}{\mu_0(1+\kappa)E(x)}, \quad (17)$$

where the lateral electric field in the channel $E(x)$ is an accurate solution of Eq.(5).

## III. APPLICATIONS

### A. Inverse logarithmic slope and transconductance

Another important small-signal characteristics of the field-effect devices is the inverse logarithmic slope $S \equiv I_D(dV_G/dI_D)$, which is a measure of the gate control over drain current, relating with the transconductance $g_m = dI_D/dV_G$ as follows

$$S = I_D(dV_G/dI_D) = I_D/g_m. \quad (18)$$

This relation makes $S$ an extremely important parameter in the well-known $g_m/I_D$ sizing methodology in the analog CMOS circuits [10]. Fig. 1 illustrates the interrelation between $g_m$ and $S$ in (18) in a lucid graphic form.

Provided a weak dependence of the carrier's mobility on gate voltage $V_G$, the logarithmic slope can be calculated as a function of the gate bias over many decades of drain current in the following way

$$S \cong Q_C \frac{dV_G}{dQ_C} = \varphi_D\left(1 + \frac{C_D + C_{IT} + C_Q}{C_{OX}}\right) =$$
$$= \varphi_D\left(1 + \frac{C_D + C_{IT}}{C_{OX}}\right) + \frac{Q_C}{C_{OX}} \equiv m\varphi_D + \frac{Q_C}{C_{OX}}, \quad (19)$$

where $m$ is often referred to as an ideality factor. It corresponds to the well-known subthreshold slope measured in Volts per decade of the gate voltage

$$SS = S\ln 10 \cong \varphi_T \ln 10\left[1 + (C_D + C_{IT})/C_{OX}\right] \equiv m\varphi_T \ln 10. \quad (20)$$

The inverse slope above the threshold is estimated as $S \cong (V_G - V_T)/n$ ($V_T$ is the threshold voltage), where $n = 1$ in the linear regime and $n = 2$ in the saturation.

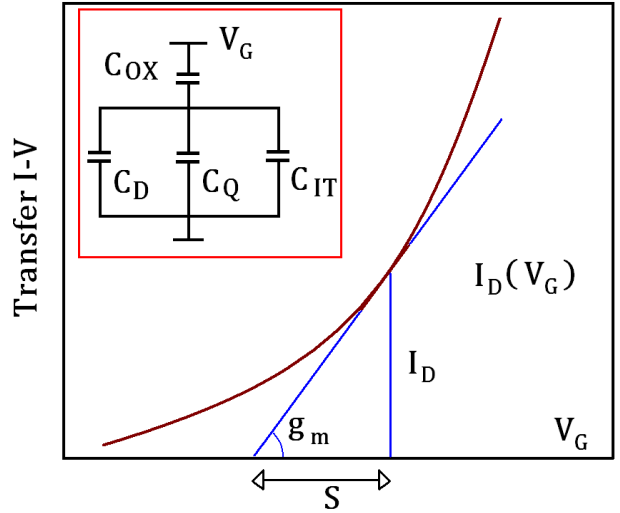

Fig. 1. Field-effect transistor transfer characteristics and graphical representation of the transconductance $g_m$ and the inverse logarithmic slope $S$.

Thus defined logarithmic slope describes in a unified way both the strong inversion and the weak inversion regions.

In contrast to the extensive tranconductance (i.e., dependent on $W/L$), the inverse logarithmic slope $S$ (as well as $V_{DSAT}$) is a thermodynamically intensive variable, i.e., independent of the channel size and shape. Comparing $V_{DSAT}$ (13) and $S$ (19), one gets

$$V_{DSAT} = \frac{2S}{1+\eta}, \quad (21)$$

where $\eta = C_D/C_{OX}$ is the substrate factor. In the strong inversion we have $V_{DSAT} \cong (V_G - V_T)/(1+\eta)$. Then, the I-V characteristics becomes

$$I_D = \frac{G_0 S}{1+\eta}\left(1 - \exp\left(-\frac{V_D}{S}(1+\eta)\right)\right). \quad (22)$$

Taking into account (18), one obtains

$$g_m = \frac{G_0}{1+\eta}\left(1 - \exp\left(-\frac{V_D}{S}(1+\eta)\right)\right) = \frac{2I_D}{(1+\eta)V_{DSAT}} \quad (23)$$

Another compact form of I-V characteristics

$$I_D = g_m\left(m\varphi_D + \frac{Q_C}{C_{OX}}\right). \quad (24)$$

### B. Subthreshold current

The MOSFET drain current (22) in the subthreshold region ($V_G < V_T$, $S \cong m\varphi_T$) becomes

$$I_D^{sub} = \frac{m}{1+\eta} \frac{D_0 Q_{C0}}{2L^2} \left(1 - \exp\left(-\frac{V_D}{m\varphi_T}(1+\eta)\right)\right) \cong$$
$$\cong \frac{m}{1+\eta} \frac{D_0 Q_T}{2L^2} e^{\frac{V_G - V_T}{m\varphi_T}} \left(1 - e^{-\frac{V_D}{m\varphi_T}(1+\eta)}\right), \quad (25)$$

where $Q_T \cong C_D \varphi_T$ is the channel charge at the threshold voltage.

### C. Gate capacitance and channel capacitance

The low-frequency gate capacitance can be defined as a derivative of the total gate charge with respect to the gate voltage

$$C_G = \left(\frac{\partial Q_G}{\partial V_G}\right) = \frac{C_Q + C_D + C_{IT}}{1 + \frac{C_Q + C_D + C_{IT}}{C_{OX}}}$$
$$= \left(\frac{1}{C_{OX}} + \frac{1}{C_Q + C_{IT} + C_D}\right)^{-1} = \frac{C_{OX}}{1+\kappa}\left(1 + \frac{C_D}{C_Q + C_{IT}}\right). \quad (26)$$

We define the channel capacitance as follows [11]

$$C_{CH} = \left(\frac{\partial Q_C}{\partial V_G}\right) = \frac{C_Q}{1 + \frac{C_Q + C_D + C_{IT}}{C_{OX}}} =$$
$$= \frac{C_{OX}}{1 + \frac{C_{OX} + C_{IT} + C_D}{C_Q}} = \frac{C_{OX}}{1+\kappa}\left(\frac{1}{1 + C_{IT}/C_Q}\right) \quad (27)$$

Equation one can get the useful relations

$$C_{CH} = \left(\frac{\kappa}{1+\kappa}\right)\frac{C_Q}{1+\eta} = \frac{2Q_C}{(1+\eta)V_{DSAT}},$$
$$= \frac{Q_C}{S} = \frac{Q_C}{\varphi_D m + Q_C/C_{OX}}, \quad (28)$$

$$\frac{Q_C}{1+\eta} = \frac{C_{CH} V_{DSAT}}{2}. \quad (29)$$

In view of (26) and (27), the gate and the channel capacitances are related with each other

$$\frac{C_G}{C_{CH}} = 1 + \frac{C_{IT} + C_D}{C_Q}. \quad (30)$$

As can be seen directly from (30), the interface traps and depletion layer impair the gate control over the channel charge.

### D. Transit time and field effect mobility

The transit time (see, equations (15-17)) can be expressed as a function of the channel capacitance and transconductance

$$\tau_{TT} = \frac{Q_C}{g_m S} = \frac{C_{CH}}{g_m} \quad (31)$$

Note that all variables are assumed to be non-linear functions of $V_G$ and $V_D$. The low-field transconductance $g_{m0}$ is closely related with the field-effect mobility $\mu_{FE}$

$$g_{m0} = \left(\frac{\partial I_D}{\partial V_G}\right)_{V_D \to 0} = \frac{\mu_0 C_{CH} V_D}{L^2} \equiv \frac{\mu_{FE} C_{OX} V_D}{L^2}. \quad (32)$$

Then, using an operational definition of $\mu_{FE}$, we got an explicit dependence of $\mu_{FE}$ on the microscopic mobility $\mu_0$ and the small-signal capacitances

$$\mu_{FE} \equiv \frac{g_{m0} L^2}{C_{OX} V_D} = \mu_0 \frac{C_{CH}}{C_{OX}} = \frac{\mu_0}{1 + \frac{C_{OX} + C_D + C_{IT}}{C_Q}}. \quad (33)$$

The transit time in the linear regime is dependent only on microscopic mobility, and not on parasitic capacitances

$$\tau_{TT} = \frac{C_{CH}}{g_{m0}} = \frac{L^2}{\mu_{FE} V_D} \frac{C_{CH}}{C_{OX}} = \frac{L^2}{\mu_0 V_D}. \quad (34)$$

### E. Delay time and cut-off frequency in digital and analog circuits

The extrinsic cut-off frequency $\omega_T$ in analog circuits is essentially determined by an external load capacitance $C_L$ and parasitic interface trap and depletion layer capacitances

$$\omega_T = \frac{g_m}{C_G + C_L} = \frac{g_m}{C_{CH}} \frac{C_{CH}}{C_G + C_L} =$$
$$= \frac{1}{\tau_{TT}} \frac{1}{1 + \frac{C_{IT} + C_D + C_L}{C_Q}} = \frac{1}{\tau_{TT}} \frac{1}{1 + \frac{C_{IT} + C_D + C_L}{Q_C}\varphi_D} = \quad (35)$$
$$= \left[\tau_{TT} + \frac{C_{IT} + C_D + C_L}{I_D}\varphi_D\right]^{-1}.$$

$$\omega_T^{-1} = \tau_{TT} + \frac{C_{IT} + C_D}{I_D}\varphi_D + \frac{C_L}{I_D}\varphi_D. \quad (36)$$

This expression is very similar to the extrinsic cutoff frequency for bipolar transistors [12].

Further, the node delay time metrics in digital circuits can be defined as follows [13]

$$\tau_D \equiv \frac{(C_L + C_G) V_{DD}}{I_{ON}}, \quad (37)$$

where $C_L$ is the load capacitance. Then, it could be estimated accurately

$$\tau_D = \frac{C_L + C_G}{g_m} \frac{V_{DD}}{S} = \tau_{TT} \frac{C_L + C_G}{C_{CH}} \frac{V_{DD}}{S} =$$
$$= \left(\tau_{TT} + \frac{C_L + C_D + C_{IT}}{I_{ON}}\varphi_D\right)\frac{V_{DD}}{S}. \quad (38)$$

Thus, the extrinsic delay time composed of intrinsic delay, intrinsic parasitic delay and the extrinsic parasitic delay.

This point shows a complete consistency in the description of the current and capacitance parameters in the context of our approach, which is impossible without a use of explicit quantum capacitance consideration.

## IV. Discussion and conclusion

We have developed a physical approach for construction of I-V characteristics in the field-effect transistors based on a consistent solution of continuity equation in diffusion-drift approximation. It was first shown that the quantum capacitance has to be an integral part of the FET theory that allows a continuous and consistent description of I-V characteristics. The fundamental channel quantum capacitance and parasitic interface trap density spectrum have been accurately and consistently implemented into the model structure.

The simulated FET I-V characteristics were shown can be represented accurately in different equivalent forms as functions of observable differential and large-signal parameters.

A key point is that the total channel charge is considered as a basic variable. This allows considering the geometric short-channel effects as a separate engineering task, bound by electrostatics of the specific device configuration.

Another crucial short-channel effect, the carrier velocity saturation, requires generally an accurate solution of the current continuity equation. The problem is the dependence of the drift velocity on the channel electric filed is generally unknown and often approximated phenomenologically in an inconvenient piecewise form [14]. This makes the exact analytical solution of the continuity equation meaningless since the choice of the mobility dependence approximation $\mu_0(E)$ strongly affects the functional form of the continuity equation solution [15]. An alternative phenomenological approach was proposed in [8, 16] where the electrostatic saturation voltage $V_{DSAT}$ in a basic equation (11) is replaced by a generalized saturation voltage $V_{SAT}$, defined as

$$V_{SAT} = \frac{2v_{\max}L}{\mu_0}\tanh\left(\frac{\mu_0 V_{DSAT}}{2v_{\max}L}\right), \tag{39}$$

that corresponds to $I_{DSAT} \cong (1/2)G_0 V_{DSAT} \propto (V_G - V_T)^2$ for the long-channel devices (square-law approximation), and the saturation current for the short channels $I_{DSAT} \cong (1/2)G_0 V_{SAT} \cong v_{\max}Q_C/L$ is determined by the maximum drift velocity $v_{\max}$ of the channel carriers. This approach was validated on the extensive experimental data (see, e.g., [17]).

We believe that being based on the fairly general physical principles, the described approach can be the basis for the development of the universal compact models of the field-effect transistors of different types.